\documentclass[twocolumn,showpacs,preprintnumbers]{revtex4}
\usepackage{amssymb}
\usepackage{graphicx,epsf}

\begin{document}
\draft
\title{Decay process accelerated by tunneling in its very early stage}
\author{T. Koide$^a$ and F.M. Toyama$^b$ }
\address{$^a$Yukawa Institute for Theoretical Physics, Kyoto University, Kyoto
606-8502, Japan}
\address{$^b$Department of Communication and Information Sciences, Kyoto Sangyo
University, Kyoto 603-8555, Japan}
\date{\today}

\begin{abstract}
We examine a fast decay process that arises in the transition period between 
the Gaussian and exponential decay processes in quantum decay systems. 
It is usually expected that the decay is decelerated by a confinement 
potential barrier.
However, we find a case where the decay in the transition period is 
accelerated by tunneling through a confinement potential barrier.
We show that the acceleration gives rise to an appreciable effect on
the time evolution of the nonescape probability of the decay system.  
\end{abstract}

\pacs{03.65.-w,03.65.Xp}

\maketitle

\noindent

Quantum decay processes such as the nuclear alpha decay are usually well
described by means of the exponential decay law; see, for example,
\cite{ref:Gamow,ref:We-Wi,ref:Br-Wi}.
Theoretically, however, deviations 
from the exponential law are expected in the beginning and also
toward the end of a decay process \cite{ref:review}. It is understood that
the quantum decay process in general proceeds through three different
stages; initial, intermediate and final. 
The initial stage is characterized 
by the Gaussian law, the intermediate stage by the exponential law, and the
final stage by the power law. 
The decay speeds in the initial and final stages are 
smaller than that in the intermediate stage.
In particular, the slow decay process in the Gaussian period leads to 
the possibility of the quantum Zeno effect \cite{ref:review}, in which
the decay process is decelerated by repeated measurements.

Many years ago, 
in his one-dimensional model analysis of a decaying system, 
Winter found that the speed of the decay process exhibits an irregular
behavior in the transition period  between the Gaussian and 
exponential periods \cite{ref:Win}. 
Very recently Dicus et al. reexamined the same system, which consists 
of a particle which is initially confined in a region and leaks out by
tunneling through a {\em delta-function potential barrier} \cite{ref:DICUS}.
In the irregular decay process, 
the decay speed can be larger than that in the exponential region.
Such fast decay in the transition period 
is interesting in the sense that it may 
give rise to acceleration of the decay process by repeated measurements, 
which is the so-called anti- (inverse-) Zeno effect \cite{ref:IQZE,ref:Fisc}.

The purpose of this report is to investigate the details of
the fast decay process in the transition period between the Gaussian and 
exponential periods \cite{footnote:0}.
In particular, we explore how the decay process in the transition period
depends on the strength of the confinement potential barrier
with a {\em finite potential width}.
It is usually expected that the decay is suppressed as the 
confinement becomes stronger because the potential barrier 
suppresses the time evolution of the wave function.
However, we will illustrate a case where the decay in the transition period 
is accelerated by tunneling through a confinement potential barrier.
Further, we show that the acceleration gives rise to an appreciable effect on
the time evolution of the nonescape probability of the decay system.

We consider a model in one dimension with a potential that consists of an
infinite repulsive wall at $x=0$ and a repulsive square barrier at $1 < x <
1+w$, 
\begin{eqnarray}
V(x) = \left\{
\begin{array}{cl}
0 & {\rm for}~~0 < x < 1,\\ 
h & {\rm for}~~1 < x < 1+w, \\ 
0 & {\rm for}~~1+w < x,
\end{array}
\right.
\end{eqnarray}
where $h$ and $w$ are the height and width of the potential, respectively.

We assume that a particle of mass $m$ is initially confined within the
potential barrier and it leaks out in time. The time-evolution of the system
is determined by the time-dependent Schr\"odinger equation 
\begin{eqnarray}
i\frac{\partial \phi (x,t)}{\partial t} = H \phi(x,t),  ~~~H =  -\frac{\partial^2}{\partial
x^2} + V(x)  \label{eqn:TDSE}
\end{eqnarray}
where the units are such that $\hbar=1$ and $2m=1$. We solve Eq. (\ref
{eqn:TDSE}) numerically using the implicit solution method\cite{ref:NC},
with the unitary time-evolution operator $U\equiv(1-iH\delta
t/2)/(1+iH\delta t/2)$, where $\delta t$ is the {time mesh}. For the range
of $x$, we take [0,500]. In order to suppress reflection of the wave
function at the boundary ($x=500$) of the model space, we assume an
imaginary potential at $x \ge 490$.

For the initial wave function of the particle which is confined in the
region of $0\leq x \leq 1$, we assume 
\begin{equation}
\phi(x,0) = \sqrt{2}\sin \pi x.  \label{eqn:IWF}
\end{equation}
It is understood that $\phi(x,0)=0$ for $x>1$. The wave function leaks out
in time by tunneling through the potential barrier $V(x)$.
The energy expectation value $<H>$ at $t=0$ is $\pi^2$.
In numerical illustrations we choose the height of the potential
barrier such that $h>\pi^2$. 
In this sense we regard the process as a tunneling decay process.

We define probability $P(a,t)$ for the particle being in the interval $0\leq
x \leq a$ at $t$ by 
\begin{eqnarray}
P(a,t) = \int^{a}_{0}dx |\phi(x,t)|^2.
\end{eqnarray}
If we set $a=1+w$, $P(1+w,t)$ represents the probability that the particle
is confined by the potential barrier. For $w$ we consider a few different
values. In order to compare the results for different values of $w$, we set $
a$ such that $a>1+w_{max}$. Throughout this paper, we take $a=4$. The $P(a,t)
$ is a kind of the nonescape probability, which is the probability that the
particle has not escaped from the potential by time $t$
\cite{ref:F-T,ref:GC-M-M}.

Next we introduce function $g(a,t)$ defined by 
\begin{eqnarray}
g(a,t) = \frac{d P(a,t)/dt}{P(a,t)}.
\end{eqnarray}
If probability $P(a,t)$ decays exponentially, that is, if $P(a,t) \propto
e^{-\gamma t}$, then $g(a,t)$ is independent of time, 
\begin{eqnarray*}
g(a,t) = -\gamma.
\end{eqnarray*}
If $P(a,t)$ obeys the Gaussian decay law, that is, if $P \propto
e^{-t^2/\tau}$, $g(a,t)$ is proportional to $t$: 
\begin{eqnarray}
g(a,t) = -2t/\tau.
\end{eqnarray}
Thus, from the $t$ dependence of $g(a,t)$ we can see how well $P(a,t)$ obeys
the exponential law or the Gaussian law.

Figure 1 shows the nonescape probability $P(a,t)$, where the potential 
height and width are taken as $h=10$ and $w=0.6$, respectively.
In the initial period $0\leq t\lesssim 0.3$, the decay is extremely slow.
The period should correspond to the Gaussian decay process.
In the period of $t\gtrsim 2$, 
the system is subject to the exponential decay law.
Between the two stages, there is a period ($0.3\lesssim t\lesssim 2$) 
in which the decay process is neither Gaussian nor exponential. 
This is the transition period. 
In the period the decay of the nonescape probability 
seems to be much faster than that
in the exponential period.

\begin{figure}[tbp]
\begin{center}
\leavevmode \epsfxsize=8cm \epsfbox{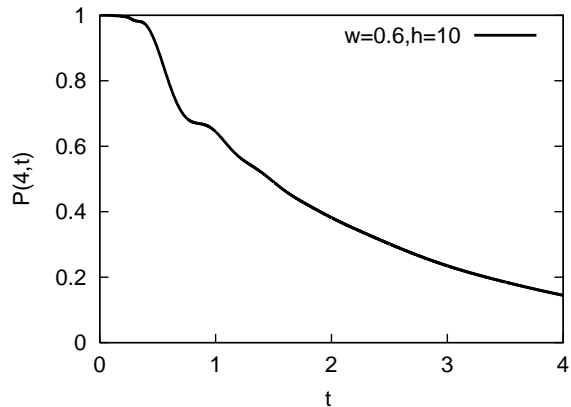}
\end{center}
\caption{ The time-evolution of the nonescape probability $P(4,t)$ for $w=0.6
$ and $h=10$. The units are such that $\hbar=1$ and $2m=1$. }
\end{figure}

\begin{figure}[tbp]
\begin{center}
\leavevmode \epsfxsize=8cm \epsfbox{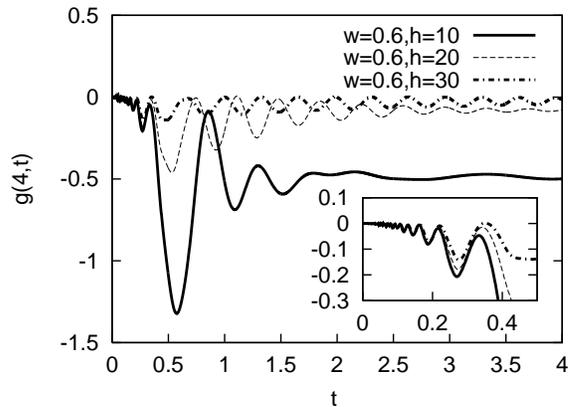}
\end{center}
\caption{ The time-evolution of $g(4,t)$ for a fixed width $w=0.6$. The
solid, dashed and dot-dashed lines show $g(4,t)$ for the heights 
$h=10$, $20$ and $30$, respectively. The units are such that $\hbar=1$ 
and $2m=1$. }
\end{figure}

The function $g(4,t)$ calculated from $P(4,t)$ of Fig. 1 
is shown by the solid line in Fig.~2.
In the period of $t\gtrsim 2$, $g(4,t)$ is almost constant, which means that
the decay process is exponential. 
In the period of $0\leq t\lesssim 0.3$, 
$g(4,t)$ is not exactly proportional to $t$, i.e.,
the decay process in the initial stage slightly deviates 
from the Gaussian decay law.
However, the decay speed in this period is still 
smaller than that in the exponential period.
Corresponding to the rapid decay process seen in $P(t)$ of Fig. 1, 
the maximum of $|g(4,t)|$ is obtained in the transition period 
($t \sim 0.6$).
The $g(4,t)$ starts from zero at $t=0$.
Thus, the quantum Zeno effect is possible when we repeat 
measurements {\ with} a sufficiently small time interval.
On the other hand, 
the anti-Zeno effect is possible by repeated measurements
only when the net decay rate of the fast decay process in the transition 
period is large compared to that of the slow decay process in the Gaussian
period (see footnote \cite{footnote:1}).

\begin{figure}[tbp]
\begin{center}
\textbf{\leavevmode \epsfxsize=8cm \epsfbox{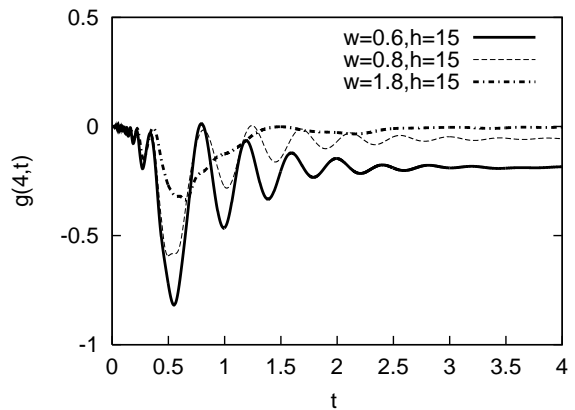} }
\end{center}
\caption{ The time-evolution of $g(4,t)$ with a fixed height $h=15$. The
solid, dashed and dot-dashed lines are the $g(4,t)$ for $w=0.6$, $0.8$ and $
1.8$, respectively. The units are such that $\hbar=1$ and $2m=1$. }
\end{figure}

In order to see how the decay process in the transition period
depends on the strength of the potential barrier,
we examine $g(4,t)$ for various potential heights and widths.
The dashed and dotted-dashed lines in Fig. 2 show 
the $g(4,t)$ for $h=20$ and $h=30$, respectively with a fixed width 
$w=0.6$.
The decay process in the Gaussian period $0\leq t\lesssim 0.3$ 
does not depend strongly on the potential height.
This is because in the initial stage the higher energy components of the 
initial wave function relative to the potential height contribute mainly to 
the decay of the system. 
On the other hand, the decay speed in the exponential period becomes 
much smaller as the confinement becomes stronger.
The fast decay in the transition period depends strongly on 
the potential height.
It tends to be suppressed as the confinement becomes stronger.

In Fig. 3, we show $g(4,t)$ for
various potential widths and a fixed height $h=15$. 
The decay process in the Gaussian period $0\leq t\lesssim 0.3$ 
does not seem to depend strongly on the potential width. 
On the other hand, 
the decay rate of the exponential period becomes 
much smaller as the potential width becomes broader. 
This is due to the increase of the confinement strength.
The decay speed in the transition period becomes 
larger as the potential width becomes narrower.

As we have shown, 
the speed of the fast decay process becomes smaller as 
the potential barrier becomes stronger.
Thus, we might guess that the fastest decay 
will be obtained in the decay process with no potential barrier. 
However, as we will show in the following, 
potential barriers with appropriate 
widths and heights can accelerate the decay process.

\begin{figure}[tbp]
\begin{center}
\textbf{\leavevmode \epsfxsize=8cm \epsfbox{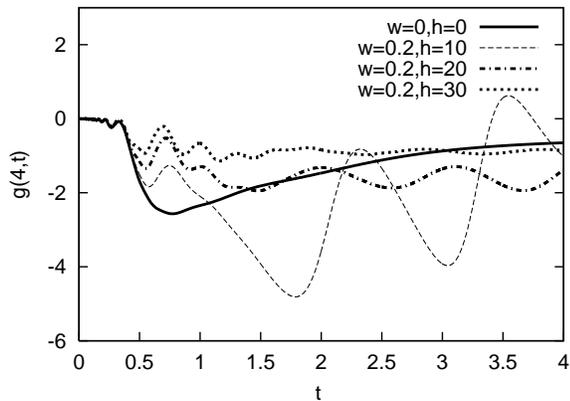} }
\end{center}
\caption{ The time-evolution of $g(4,t)$ for a fixed width $w=0.2$. The
dashed, dot-dashed and dotted lines are the $g(4,t)$ for the widths 
$h=10, 20$ and $30$, respectively. The solid line shows the $g(4,t)$ for no
potential barrier. The units are such that $\hbar=1$ and $2m=1$. }
\end{figure}

In Fig.~4, the solid line shows $g(4,t)$ calculated with no potential barrier. 
The dashed, dot-dashed and dotted lines 
exhibit $g(4,t)$ calculated with the potential 
barriers with $h=10, 20$ and $30$ and fixed width $w=0.2$, respectively.
It should be noted that $w=0.2$ is much thinner than those used in 
Figs.~2 and 3, and 
therefore, in this case, the confinement is very weak compared 
with that in Figs.~2 and 3.
In the Gaussian period, the potential dependence of the decay speed
is not very appreciable.
However, in the transition region, 
the decay speed becomes faster as the potential height becomes lower.
For $h=10$, the maximum decay speed at $t \sim 1.8$ 
exceeds that for no potential barrier at $t \sim 0.7$.

As shown in Fig.~5, such an acceleration of the decay speed by tunneling 
gives rise to an appreciable difference in the time evolution of
$P(4,t)$.
The solid and dashed lines are the $P(4,t)$ for no potential barrier and for 
the potential barrier with $w=0.2$ and $h=10$, respectively.
The nonescape probability for $h=10$ becomes smaller 
than that for no potential barrier at $t \sim 1.5$.
At this time the residual nonescape probability is still about ten percent. 
In this sense, the effect of this acceleration cannot be ignored.
On the other hand, for $h=20$, at the time region in which the nonescape
probability becomes smaller than that for no potential barrier, 
the residual nonescape probability is negligibly small.

\begin{figure}[tbp]
\begin{center}
\textbf{\leavevmode \epsfxsize=8cm \epsfbox{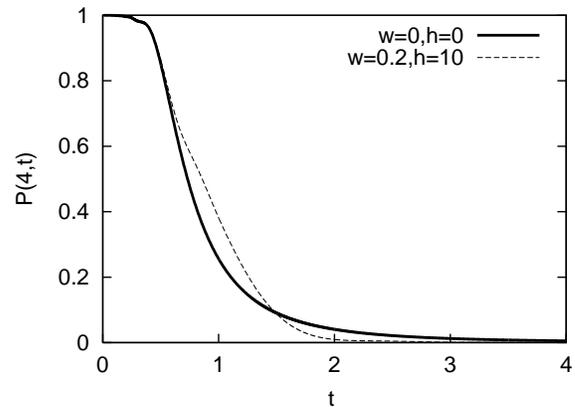} }
\end{center}
\caption{ The time evolution of the nonescape probabilities $P(4, t)$. The
solid and dashed lines are the $P(4, t)$ for no potential barrier and for
the potential barrier with $w=0.2$ and $h=10$, respectively. The units are
such that $\hbar=1$ and $2m=1$. }
\end{figure}

The fluctuations in the behaviors 
of $g(4,t)$ indicate that the decay processes are still 
in the transition period 
from the Gaussian to exponential period.
However, as shown in Fig. 5, the decay process has been almost 
completed before $t=4$.
Therefore, even if the decay process proceeds to the exponential period 
eventually, the exponential decay has no importance in this case. 
The stability of quantum system is usually characterized by the 
magnitude of the imaginary part of the pole 
that gives the inverse of the lifetime in 
the exponential period \cite{ref:review,ref:Win,ref:DICUS}. 
The result that we have shown implies that such a pole analysis may not 
be effective for highly unstable systems \cite{ref:ModOpt}.

In Fig.~4, we have seen 
that the acceleration by tunneling can be obtained 
for $h \lesssim 10$.
Next, we investigate the acceleration with a fixed height.
We examine the time evolution of $g(4,t)$ for $h=10$ \cite{footnote:2}.
The dashed, dot-dashed and dotted lines in Fig. 6 
exhibit $g(4,t)$ calculated for the potential 
barriers with $w=0.2, 0.4$ and $0.6$ and fixed height $h=10$, respectively.
One can see that the decay rate becomes larger for thinner 
potential widths. 
For $w=0.2$, the maximum decay speed exceeds that for no potential barrier.
Thus, one sees that the acceleration of the decay speed by tunneling 
can be obtained when the strength of the confinement by the potential barrier
is sufficiently weak.
In our illustrations, for $h \lesssim 10$ and $w \lesssim 0.2$, 
the accelerations are remarkable.

Finally, we mention that the $g(4,t)$ represented 
by dotted line in Fig.~4 ($w=0.2$ and $h=10$) 
takes positive values around $t\sim 3.5$.
Recall that the probability current $j(a,t)$ is related to 
the nonescape probability $P(a,t)$ by 
$j(a,t)=-dP(a,t)/dt=-g(a,t)P(a,t)$.
This means that, if $g(a,t)$ is positive, $j(a,t)$ is negative. 
However, in \cite{ref:Win,ref:NTD} 
the negative currents were obtained at very late time 
region after the exponential decay period. 
Our result implies that in a highly unstable quantum system the negative 
current can occur even at the initial stage after the Gaussian period.

\begin{figure}[tbp]
\begin{center}
\textbf{\leavevmode \epsfxsize=8cm \epsfbox{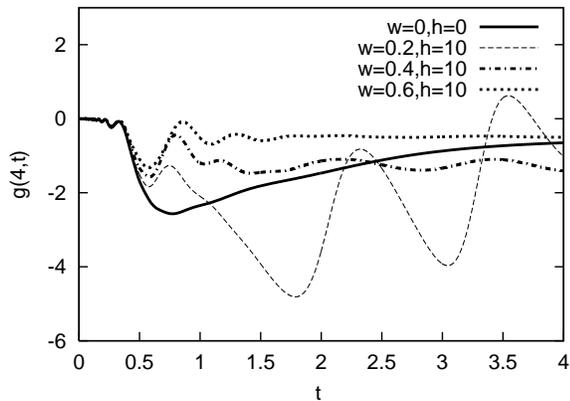} }
\end{center}
\caption{ The time-evolution of $g(4,t)$ with a fixed height $h=10$. The
dashed, dot-dashed and dotted lines are the $g(4,t)$ for the heights
$w=0.2, 0.4$ and $0.6$, respectively. The solid line shows the $g(4,t)$ for no
potential barrier. The units are such that $\hbar=1$ and $2m=1$. }
\end{figure}

We have investigated the fast decay process in the transition
period between the Gaussian and exponential decay processes.
In most cases of the tunneling process,
the decay speed becomes smaller as the potential barrier becomes 
stronger.
As a special case,
we have found that the fast decay process can be remarkably accelerated
by tunneling through potential barriers 
with appropriately small widths and heights.
A detailed analysis of the acceleration 
of the fast decay process by tunneling is a future project.

\textbf{\ \newline
}

\textbf{Acknowledgment\newline
}

We would like to thank Professor Y. Nogami for useful discussions and
comments.  This work was supported in part by the Minstry of Education,
Culture, Science, Sports and Technology of Japan.

\textbf{\ \newline
}


\begin{thebibliography}{99}

\bibitem{ref:Gamow}
G.~Gamow, Z.~Phys. {\bf 51}, 204 (1928); {\bf 52}, 510 (1928);
E.~U.~Condon and R.~W.~Gurney, Nature {\bf 112}, 439 (1928);
Phys.~Rev. {\bf 33}, 127 (1929).
%
\bibitem{ref:We-Wi}
V.~Weisskopf and E.~P.~Wigner, Z.~Phys. {\bf 63}, 54 (1930);
65 (1930).
%
\bibitem{ref:Br-Wi}
G.~Breit and E.~P.~Wigner, Phys.~Rev. {\bf 49}, 519 (1936).
%
\bibitem{ref:review}
For review articles, see
L.~Fonda, G.~C.~Ghirardi and A.~Rimini,
Rep.~Prog.~Phys. {\bf 41}, 587 (1978);
H.~Nakazato, M.~Namiki and S.~Pascazio,
Int.~J.~Mod.~Phys. B {\bf 10}, 247 (1996).
%
\bibitem{ref:Win}
R.~G.~Winter, Phys.~Rev.~{\bf 123}, 1503 (1961).
%
\bibitem{ref:DICUS}
D.~A.~Dicus, Wayne~W.~Repko, Roy~F.~Schwitters, and Todd~M.~Tinsley,
Phys.~Rev. A {\bf 65}, 032116 (2002).
%
\bibitem{ref:IQZE}
W.~C.~Schieve, L.~P.~Horwitz and J.~Levitan, Phys.~Lett. A {\bf 136},
264 (1989);
A.~G.~Kofman and G.~Kurizki, Nature(London) {\bf 405}, 546 (2000);
P.~Facchi, H.~Nakazato and S.~Pascazio,
Phys.~Rev.~Lett. {\bf 86}, 2699 (2001) and referencees therein.
%
\bibitem{ref:Fisc}
%As a recent experimental result, 
M.~C.~Fischer, B.~Guti\'errez-Medina and M.~G.~Raizen,
Phys.~Rev.~Lett. {\bf 87}, 040402 (2001).
%
\bibitem{footnote:0}
The transition region between the exponential and power periods 
has been examined; 
see W.~van~Dijk and Y.~Nogami, Phys.~Rev.~C{\bf 65}, 024608 (2002).
%
\bibitem{ref:NC}
A.~Goldberg,~H.~M.~Schey and J.~Schwartz,
Am.~J.~Phys. {\bf 65}, 177 (1967).
%
\bibitem{ref:F-T}
R.~B.~Frey and E.~Thiele, J.~Chem.~Phys. {\bf 48}, 3240 (1968).
%
\bibitem{ref:GC-M-M}
G.~Garc\' \i a-Calder\'on, J.~L.~Mateos and M.~Moshinsky,
Phys.~Rev.~Lett. {\bf 74}, 337 (1995);Ann.~Phys. (NY) {\bf 249},
430 (1996).
%
\bibitem{footnote:1}
In order to observe the anti-Zeno effect, 
we have to repeat 
the measurements with a time interval $\tau$ that satisfies the condition 
$F(\tau) = \int^{\tau}_{0}dt (g(4,t)+\gamma_{exp}) < 0$, 
where $\gamma_{exp}$ is a decay rate in the exponential region.
%
\bibitem{ref:ModOpt}
A.~G.~Kofman,~G.Kurizki and B.~Sherman, 
J.~Mod.~Opt. {\bf 41}, 353 (1994).
%
\bibitem{footnote:2}
Notice that we consider the tunneling phenomena, so 
we do not investigate the lower height than $H=10$.
%
\bibitem{ref:NTD}
Y.~Nogami, F.~M.~Toyama and W.~van Dijk,
Phys.~Lett. A {\bf 270}, 279 (2000).
%
\end{thebibliography}
\end{document}